# The molybdenum-titanium phase diagram evaluated from *ab-initio* calculations

S. Barzilai[1], C. Toher[2], S. Curtarolo[2], O. Levy[2,3]
[1] Department of Materials Science, NRCN, P.O.Box 9001, Beer-Sheva 84190, Israel.
[2] Department of Mechanical Engineering and Materials Science, Duke University, Durham, North Carolina 27708, USA
[3] Department of Physics, NRCN, P.O.Box 9001, Beer-Sheva 84190, Israel.

Abstract
The design of next generation *β*-type titanium implants requires detailed knowledge of the relevant stable and metastable phases at temperatures where metallurgical heat treatments can be performed. Recently, a standard specification for surgical implant applications was established for Mo-Ti alloys. However, the thermodynamic properties of this binary system are not well known and two conflicting descriptions of the *β*-phase stability have been presented in the literature. In this study, we use *ab-initio* calculations to investigate the Mo-Ti phase diagram. These calculations predict that the *β*-phase is stable over a wide concentration range, in qualitative agreement with one of the reported phase diagrams. In addition, they predict stoichiometric compounds, stable at temperatures below $300^0$C, which have not yet been detected by experiments. The resulting solvus, which defines the transition to the *β*-phase solid solution, therefore occurs at lower temperatures and is more complex than previously anticipated.

## 1. Introduction

The demand for permanent implants in the human body grows as people live longer and their bones weaken with age. *β*-type titanium alloys are known to be one of the best choices for biomedical applications based on their excellent biocompatibility in the human body environment, high strength, enhanced corrosion resistance and relatively low elastic moduli [1, 2, 3, 4, 5, 6, 7, 8, 9]. Titanium alloys with low Young's moduli inhibit the stress shielding effect and thus effectively avoid bone atrophy and enhance bone remodeling. They are therefore an attractive choice for replacing failed hard tissue [10, 11]. Additional properties, such as small spring back, low yielding stress and high ultimate strength, are also favorable for achieving permanent compatible deformation of the implant in a narrow space in the body [10]. Design of the next generation *β*-type titanium implants requires identification of their stable and metastable phases, in particular at relatively low temperatures where metallurgical heat treatments are implemented. This is especially important for Mo-Ti alloys, for which a standard specification has been established for surgical implant applications [12], but important gaps still exist in our knowledge of their binary phase diagram.
The Mo-Ti phase diagram has been investigated experimentally and modeled using the computational thermodynamics CALPHAD approach [13] and the Thermo-Calc software [14]. These studies led to two conflicting descriptions of the stability of the *β*-phase [15, 16]. One includes a monotonic decreasing *β-transus* temperature with increasing molybdenum content [17], while the other has a complex *solvus* with a monotectoid phase separation and a miscibility gap between two different *β* phases [18]. Fig. 1 presents those two descriptions of the Mo-Ti system.

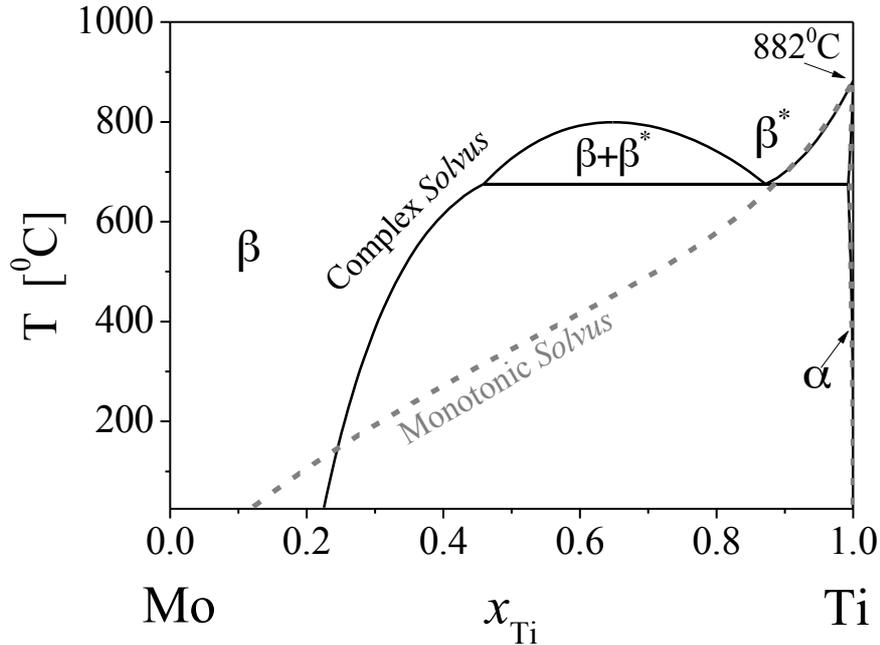

Fig. 1. Schematic description of the Mo-Ti phase diagrams assessed from the NIMS [17] (dash lines) and the Thermo-Calc [18] (solid lines) databases, respectively.

In this study we utilize density functional theory (DFT) to compute the Mo-Ti phase diagram below 900ºC. To the best of our knowledge there are no published *ab-initio* studies of this phase diagram, although formation enthalpies of the $\alpha$ and $\beta$ phases [19] and the mixing enthalpy of the $\beta$-phase [20, 21, 22] were computed. Here we employ the AFLOW high-throughput framework [23] to screen a database of ordered intermetallic structures and estimate the formation enthalpies of bcc ($\beta$) and hcp ($\alpha$) solid solutions using the special quasi-random structures (SQS) methodology [24]. The ideal expression for the configuration entropy and the quasi-harmonic Debye model for the vibrational energy are used to estimate the finite-temperature contributions to the free energy.

This analysis leads to a binary phase diagram that is qualitatively different from those previously reported for the Mo-Ti system. It includes stable compounds and a much more complex transition to the $\beta$-phase over the entire range of compositions. These differences emerge at low temperatures, where they have escaped detection due to the influence of slow kinetics and very long equilibration processes on the experimental results.

## 2. Methodology
2.1. DFT calculations

We start by screening an extensive database of 900 ordered structures via the high-throughput framework AFLOW [23]. In addition, we calculate the formation energies of the pure elements in the hcp and bcc structures and the SQS for these structures at compositions $Mo_{0.75}Ti_{0.25}$, $Mo_{0.5}Ti_{0.5}$ and $Mo_{0.25}Ti_{0.75}$. The SQS calculations employed the 16-atom unit cell structures reported in [25] and [26] for the bcc and hcp cases, respectively. All total energy calculations were carried out using the VASP software [27] within the AFLOW standard for material structure calculations [23,28], with projector augmented waves (PAW) pseudopotentials [29] and the exchange correlation functionals parameterized by Perdew, Burke, and Ernzerhof (PBE) [30] for the generalized gradient approximation (GGA). All crystal structures were fully relaxed (cell volume and shape and the basis atom coordinates inside the cell). Numerical convergence to about 1

meV/ atom was ensured by a high-energy cutoff, 30% higher than the highest energy cutoff for the pseudopotentials of the components, and dense Monkhorst-Pack meshes [31] with at least 6000 k-points per reciprocal atom. Complete information about these calculations is included in the open access AFLOW.org materials data repository [32, 33].

The calculations of the total energies of the pure elements and the relaxed hcp- and bcc-SQS were repeated using a full potential method, employing the Augmented Plane Waves + local orbitals (APW+lo) formalism as implemented in the WIEN2k code [34, 35]. In these calculations the core states treatment is fully relativistic [36] and the valence states are considered in the scalar relativistic approximation [37]. The GGA-PBE exchange correlation potential was employed, as in the VASP calculations. The radii of the muffin-tin spheres ($R_{mt}$) were 2.3 a.u. for both Mo and Ti. It was found that a basis-set size of $R_{mt}K_{max} = 11$, where $K_{max}$ represents the magnitude of the largest K vector in the wave function expansion, and a $k$-mesh of 600 points for the SQS and 3500 points for the pure elements suffices to reach an accuracy of ~$10^{-4}$ Ry in the total-energy calculations, with an energy cutoff separating core and valence states of -6 Ry.

2.2. Thermodynamic modeling

The formation enthalpy of a binary intermetallic structure is

(1) $H_F(Mo_{(1-x)}Ti_x) = H(Mo_{(1-x)}Ti_x) - (1-x)H(Mo) - xH(Ti)$

where $H(Mo_{(1-x)}Ti_x)$ is the enthalpy per atom of the intermetallic structure and $H(Mo)$ and $H(Ti)$ are the enthalpies per atom of the elements at their ground state structures, bcc-Mo and hcp-Ti. Negative formation enthalpies signify structures that are energetically favorable compared to phase separation into the elemental structures and are therefore possible candidates for stable compounds in the binary system.

The vibrational contribution to the free energy, $G_{vib}$, is estimated by the Debye model [38,39]

(2) $G_{vib} = \frac{9}{8}R\theta_D + RT\left[3\ln\left(1-\exp(-\frac{\theta_D}{T})\right) - 3\frac{T^3}{\theta_D^3}\int_0^{\theta_D/T}\frac{z^3 dz}{\exp(z)-1}\right]$

where

(3) $\theta_D(T) = \frac{\hbar}{k_B}\left(6\pi^2\sqrt{V}n\right)^{\frac{1}{3}} f(\sigma)\sqrt{\frac{B_T}{M}}$

is the Debye temperature,

(4) $f(\sigma) = \left\{3\left[2\left(\frac{2}{3}\frac{1+\sigma}{1-2\sigma}\right)^{3/2} + \left(\frac{1}{3}\frac{1+\sigma}{1-\sigma}\right)^{3/2}\right]^{-1}\right\}^{1/3}$,

n is the number of atoms in the unit cell, M is its mass and V is its volume, $\sigma$ is the Poisson ratio and $B_T$ is the bulk modulus at finite temperatures.

The free energies of the bcc and hcp solid solutions for Mo-Ti alloys are expressed by

(5) $G^\varphi(x_{Mo}, x_{Ti}, T) = x_{Mo} \cdot {}^0G^\varphi_{Mo} + x_{Ti} \cdot {}^0G^\varphi_{Ti} + {}^{mix}G^\varphi + {}^{ex}G^\varphi$

where φ represents the bcc or the hcp structures, $^0G_{Mo}$ and $^0G_{Ti}$ are the Gibbs energies of the pure elements, T is the absolute temperature, $^{mix}G^\varphi$ is the configuration energy of the ideal solution

(6) $^{mix}G^\varphi(x_{Ta}, x_{Ti}, T) = kT(x_{Ta}\ln x_{Ta} + x_{Ti}\ln x_{Ti})$

and $^{ex}G^\varphi$ is the excess energy representing the effect of non-ideality

(7) $^{ex}G^\varphi = {}^{ex}H^\varphi + {}^{ex}G^\varphi_{vib}$.

The excess enthalpy, $^{ex}H^\varphi$, is computed directly from the DFT results for the corresponding SQS and the pure element structures

(8) $^{ex}H^\varphi = H^\varphi_{MoTi} - x_{Mo} \cdot H^\varphi_{Mo} - x_{Ti} \cdot H^\varphi_{Ti}$

and the excess vibrational energy, $^{ex}G_{vib}$, is

(9) $\quad ^{ex}G^{\varphi}_{vib} = G^{\varphi}_{vib,MoTi} - x_{Mo} \cdot G^{\varphi}_{vib,Mo} - x_{Ti} \cdot G^{\varphi}_{vib,Ti}$

The electronic and the magnetic contributions to the Gibbs energy are expected to be much smaller in this system and are therefore neglected.

## 3. Results

The high-throughput screening of 900 structures in the Mo-Ti system uncovered several stable stoichiometric structures, in contrast to the current experimental data that includes no Mo-Ti compounds. Fig. 2 presents the formation enthalpies of various ordered structures and the bcc- and hcp-SQS in this system. The new predicted stable compounds are denoted on the convex hull. As mentioned in section 2, the complete information about all these structures can be found in the open access AFLOW.org materials data repository [32, 33].

The Debye temperatures of bcc-Mo, hcp-Ti and the compounds indicated in Fig. 2 are computed from the atomic volume, bulk moduli and Poisson ratios for each structure retrieved directly from the DFT calculations, as implemented in the AFLOW AEL-AGL software package [40, 41]. These results are summarized in Table 1. Fig. 3 shows the temperature dependent excess energies of the stable Mo-Ti compounds.

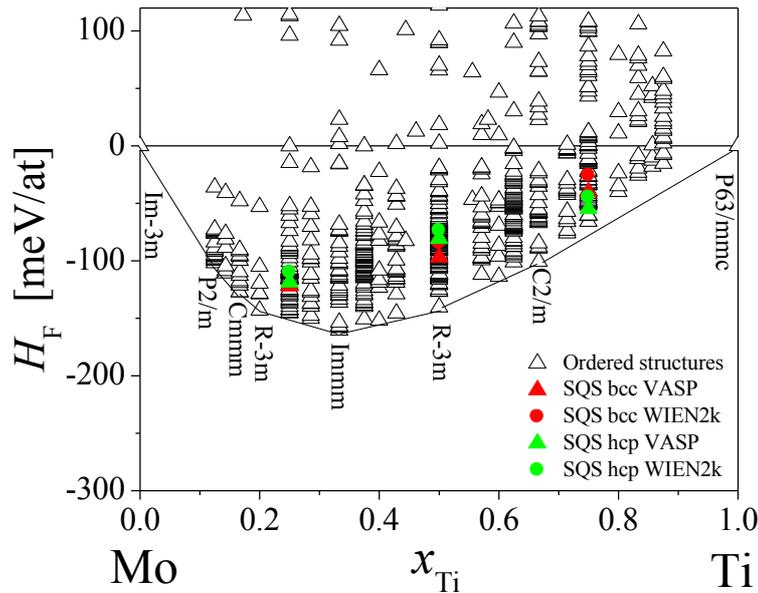

Fig. 2. Formation enthalpies of the lowest lying ordered structures and the bcc- and hcp-SQS computed with respect to bcc-Mo and hcp-Ti. Predicted compounds are denoted on the convex hull by their respective space group.

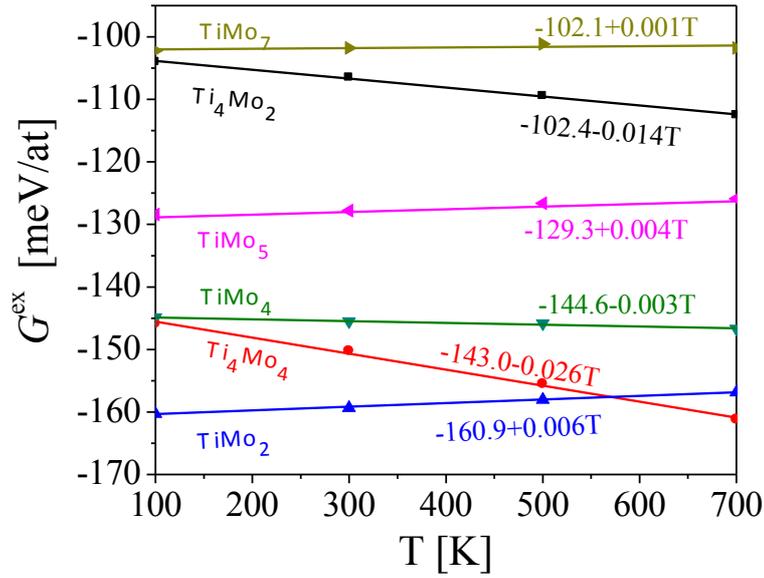

Fig. 3. The temperature dependent excess energies of the Mo-Ti compounds obtained from the *ab-initio* calculation of the zero-temperature energies and the Debye model for the vibrational energies.

The Gibbs energies of solid solutions of a specific structure are obtained by Eq. (5) from the energies of the pure elements in the same structure, and the corresponding mixing and excess energies. Here, we describe the excess free energy of the bcc and hcp structures as a function of alloy composition by fitting the excess energies of the elements and the SQS, calculated from Eq. (7), to a sub-subregular model using a Redlich-Kister polynomial of the fourth degree [42].

(10) $^{ex}G = x_{Mo}x_{Ti}\left(^{0}L_{MoTi} + ^{1}L_{MoTi}(x_{Mo} - x_{Ti}) + ^{2}L_{MoTi}(x_{Mo} - x_{Ti})^{2}\right)$.

$^{0}L$, $^{1}L$ and $^{2}L$, are the Redlich-Kister coefficients that reconstruct the interaction energy between Mo and Ti atoms in the corresponding phase. The fits for the excess enthalpies are shown in Fig. 4(a, b), with insignificant differences between the VASP and WIEN2k calculations. They represent attractive interactions between Mo and Ti for the entire range of compositions. Attractive interactions for the bcc solid solutions have been previously indicated in Refs. [19, 20, 21, 22]. However, the strong attractive interaction we obtain for the hcp solid solutions, Fig. 3(b), contrast with those published in [19] based on total energy calculations of 28 supercells, 16-atoms each, at various compositions. This difference may be related to the limited sample of structure configurations and the limited relaxation (cell volume only) of each configuration considered in [19]. Similar attractive enthalpy curves have been recently obtained for hcp solid solutions in the closely related systems Ta-Ti, Nb-Ti and V-Ti [43, 44].

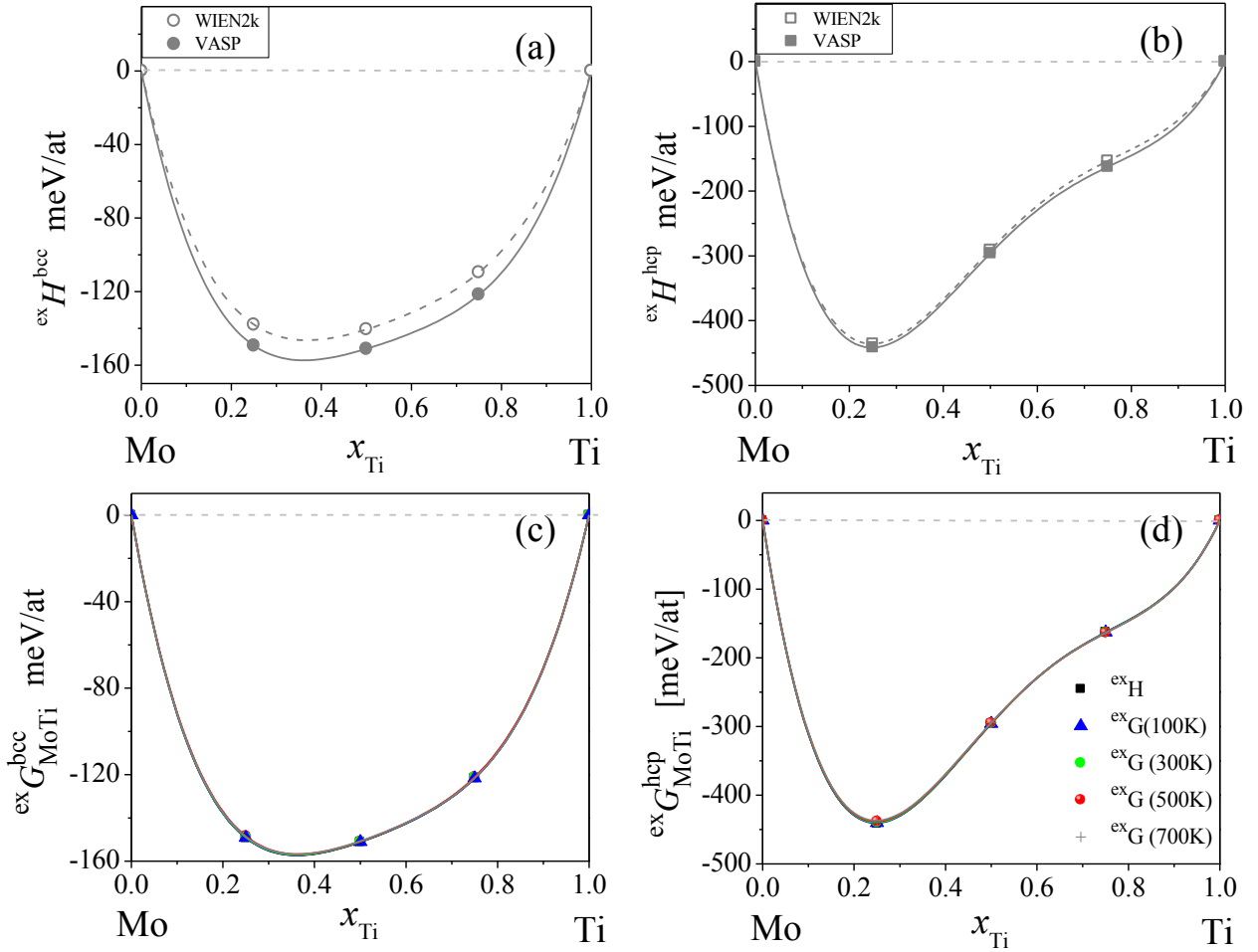

Fig. 4. Excess enthalpies (a, b) and free energies (c, d) of bcc (a, c) and hcp (b, d) Mo-Ti alloys as a function of composition. The lines are fits of the computed points to the sub-subregular model. The excess energies are attractive and their temperature dependence is negligible for both structures.

The excess Gibbs energy at finite temperatures is calculated including the mixing energy, Eq. (6), and the contribution of the vibrational energies, Eq. (9). The computed Poisson ratios, 0.308 for bcc-Mo and 0.319 for hcp-Ti, lead to scaling factors $f(\sigma)$ (Eq. (4)) of 0.742 and 0.718, respectively. These scaling factors were also used for estimating the Debye temperatures of the corresponding bcc and hcp solid solutions and to derive their temperature dependent Redlich-Kister coefficients (shown in Fig. 5). Fig. 4(c, d) shows the Redlich-Kister fits of the solid solution excess energies at selected temperatures. It turns out that their temperature dependence is negligible at the range investigated.

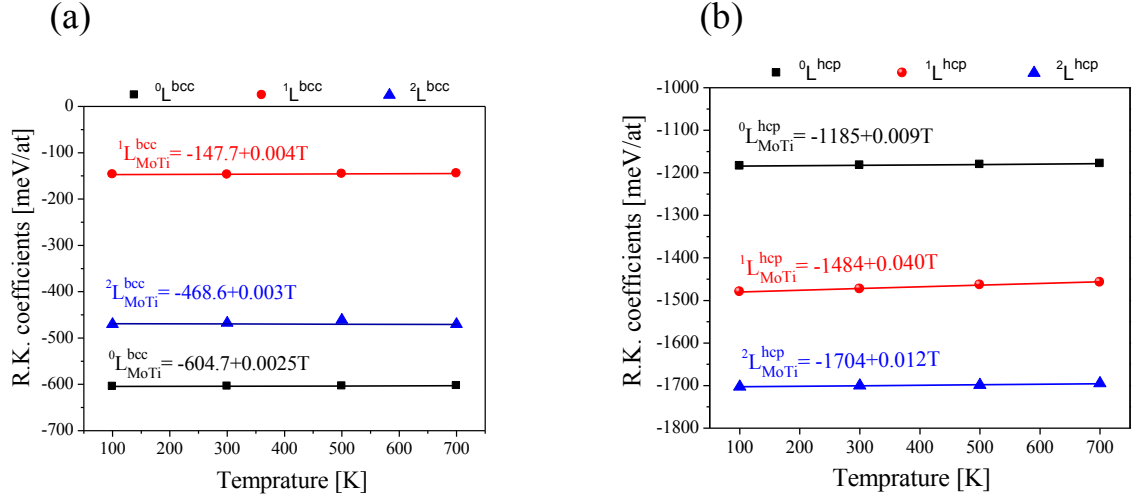

Fig. 5. The temperature dependent Redlich-Kister coefficients of the excess energies for the (a) bcc, and (b) hcp phases in the Mo-Ti system, obtained from the Debye temperatures of Table 1.

Table 1: The computed specific volumes, bulk moduli and Debye temperatures for selected structures in the Mo-Ti system.

| Phase | Composition | *Ab-initio* results | | | | | | | |
|---|---|---|---|---|---|---|---|---|---|
| | | V [Å³/at] | $f(\sigma)$ | $B_T$ [GPa] | | | $\theta_D$ [K] | | |
| | | | | 0K | 300K | 700K | 0K | 300K | 700K |
| Bcc | Ti | 17.09 | 0.742 | 104 | 101 | 97 | 406 | 401 | 392 |
| | Mo -75at%Ti | 16.64 | | 139 | 135 | 128 | 416 | 413 | 404 |
| | Mo -50at%Ti | 16.13 | | 177 | 172 | 164 | 426 | 423 | 414 |
| | Mo -25at%Ti | 15.89 | | 219 | 214 | 205 | 438 | 435 | 427 |
| | Mo | 15.86 | | 260 | 255 | 245 | 446 | 443 | 436 |
| Hcp | Ti | 17.25 | 0.718 | 110 | 107 | 99 | 406 | 401 | 391 |
| | Mo -75at%Ti | 16.61 | | 140 | 136 | 128 | 403 | 400 | 392 |
| | Mo -50at%Ti | 16.23 | | 177 | 172 | 164 | 413 | 409 | 402 |
| | Mo -25at%Ti | 15.87 | | 219 | 214 | 204 | 423 | 420 | 412 |
| | Mo | 16.18 | | 239 | 234 | 224 | 416 | 413 | 407 |
| Ordered intermetallic structures | $Mo_2Ti_4$ | 16.39 | $f(0.330) = 0.694$ | 151 | 147 | 139 | 393 | 389 | 381 |
| | $Mo_4Ti_4$ | 16.04 | $f(0.343) = 0.665$ | 176 | 172 | 163 | 381 | 378 | 370 |
| | $Mo_2Ti$ | 15.90 | $f(0.301) = 0.757$ | 205 | 200 | 191 | 443 | 439 | 431 |
| | $Mo_4Ti$ | 15.83 | $f(0.315) = 0.727$ | 228 | 223 | 214 | 431 | 428 | 421 |
| | $Mo_5Ti$ | 15.84 | $f(0.305) = 0.748$ | 235 | 230 | 220 | 447 | 443 | 436 |
| | $Mo_7Ti$ | 15.86 | $f(0.307) = 0.744$ | 241 | 236 | 226 | 446 | 443 | 435 |

It is well known that the evaluation of the thermodynamic properties of materials by the Debye model may produce values that deviate from experimental data [40, 45, 46]. To estimate the sensitivity of the excess vibrational energy to these deviations we carried out an alternative calculation, using the empirical generic scaling factors, 0.617 [45] and 0.7 [46], for the bcc and hcp structures, respectively. This calculation gave different Debye temperatures and vibrational energies but very similar excess vibrational energies for both structures. Similar insensitivity to these details of the Debye temperature calculations has been recently reported for the Ta-Ti system [43].

To compute the Gibbs energies of the bcc and hcp solutions, Eq. (5), the Gibbs energies of the pure elements, $^0G_{Mo}$ and $^0G_{Ti}$, have to be considered for each phase. It is well known that the empirical lattice stabilities of the phase structures of the elements (i.e. the energy differences between the stable phase and unstable phases), assessed, for example, by the Scientific Group Thermodata Europe (SGTE) database [47], are not consistent with first principles results for most of the transition metals [48, 49, 50, 51]. According to the SGTE, $\Delta H_{Mo}^{bcc-hcp}$=-0.118 eV/at, and $\Delta H_{Ti}^{hcp-bcc}$=-0.071 eV/at. The corresponding values obtained from our DFT calculations are $\Delta H_{Mo}^{bcc-hcp}$=-0.431 eV/at and $\Delta H_{Ti}^{hcp-bcc}$=-0.11 eV/at.

Fig. 6 shows the Gibbs energies of the compounds and the Gibbs energy curves of the solid solutions in the Mo-Ti system, computed from Eq. (5) using the SGTE values for the elements and the excess energies computed as described above with the AFLOW-AEL-AGL framework. Two stable hcp phases appear at different stoichiometries at low temperatures. Moreover, the unexpected hcp phase near Ti concentration of 0.2 survives to temperatures above $882^0$C. These features contradict the experimental data on the Mo-Ti system and are evidently incorrect. To overcome these obvious contradictions it is necessary to correct the lattice stabilities of the elements using the DFT calculated values. This means that the Gibbs energy of hcp-Mo should be increased by 0.313 eV/at compared to the SGTE value. For Ti, the known existence of a phase transition at $882^0$C allows us to derive a temperature dependent correction of $0.039-3.38\cdot10^{-5}$T eV/at. The calculations of the Gibbs energies (Fig. 7) and the phase diagram (Fig. 8) after these corrections exhibit a phase behavior much more in-line with the experimental data.

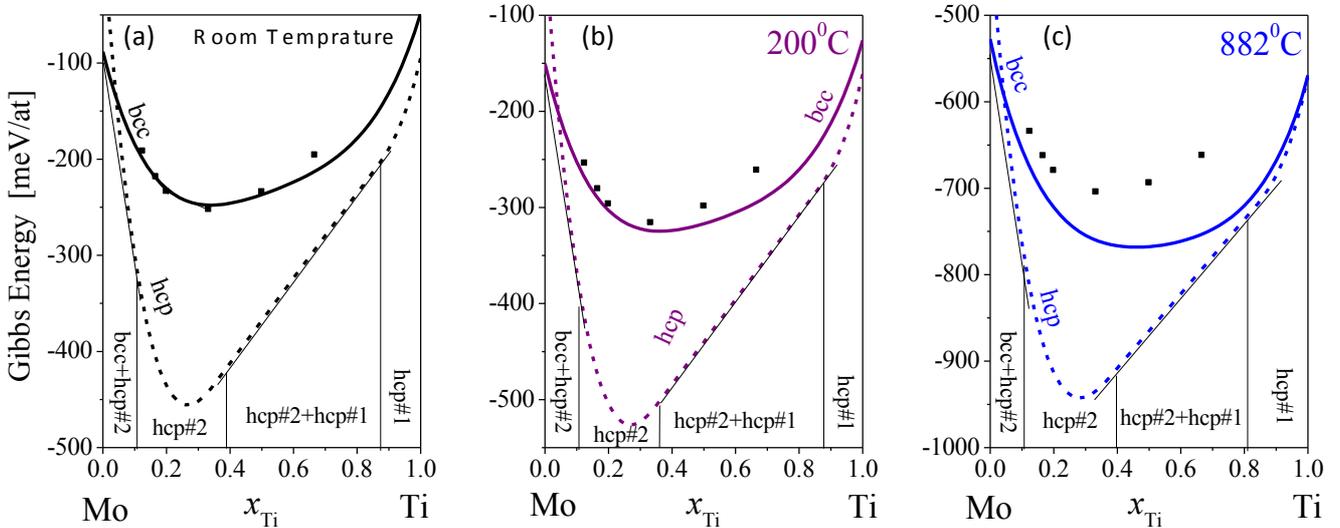

Fig. 6. Gibbs energy curves for the bcc (solid lines) and hcp (dash lines) solid solutions and the most stable ordered structures in the Mo-Ti system derived from the empirical SGTE values for the phase stabilities of the pure elements at: (a) room temperature, (b) $200^0$C and (c) $882^0$C. The tangent lines mark the stable phase convex hull of the system.

At room temperature, Fig. 7(a), the convex hull of the system includes 4 compounds in addition to a hcp phase for the Ti rich alloys and a bcc phase for the Mo rich alloys. These compounds have not been observed yet in experiments, probably due to slow kinetics at low temperatures. At higher temperatures, the Gibbs energy of the bcc phase decreases relative to the energies of the ordered structures and the stoichiometric compounds become unstable. At $200^0$C, Fig. 7(b), only one compound persists and the bcc phase stabilizes at a wide composition range. At this temperature, the convex hull includes a bcc phase up to ~38at%Ti and at the intermediate range ~65-70at%Ti, a stable compound at 50at%Ti and a hcp phase for Ti rich alloys. At 882$^0$C, just above the hcp to bcc phase transition of pure Ti, Fig. 6(c), the bcc phase becomes stable for the entire range of compositions, in agreement with the known experimental phase diagrams.

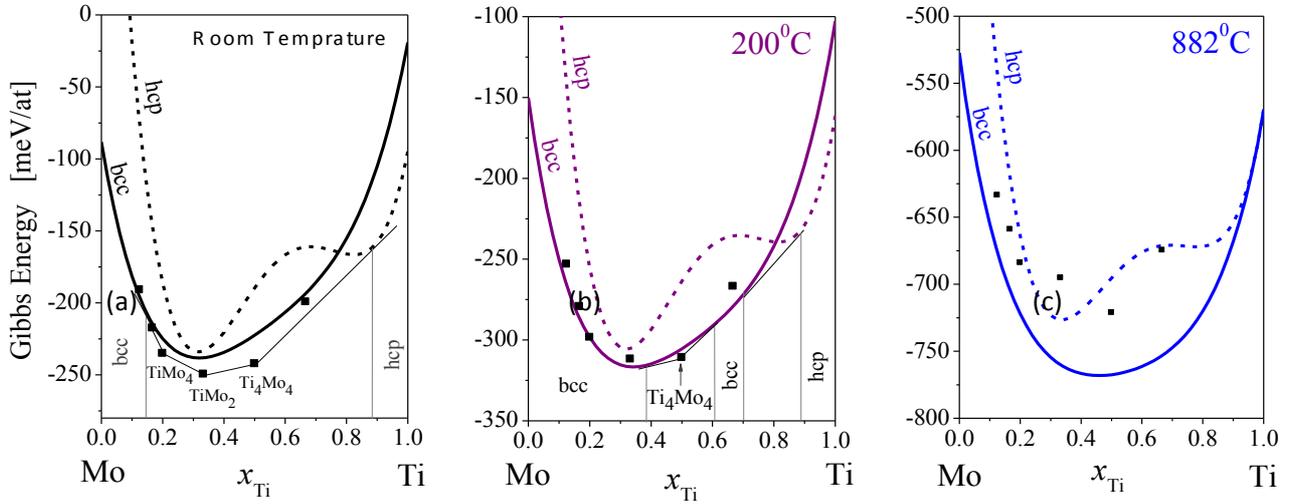

Fig. 7. Gibbs energy curves for the bcc (solid lines) and hcp (dash lines) solid solutions and the most stable ordered structures in the Mo-Ti system derived from the DFT-corrected phase stabilities of the pure elements at: (a) room temperature, (b) 200$^0$C and (c) 882$^0$C. The tangent lines mark the stable phase convex hull of the system.

Figure 8 shows the computed phase diagram of the Mo-Ti system compared to the currently reported experimental phase diagrams shown in Fig. 1. The computed phase diagram is qualitatively different from the experimental ones, exhibiting a complex solvus defined by a few compounds that transform into the disordered *β*-phase at lower temperatures than previously anticipated. It exhibits a wide stability region of the *β*-phase, similar to that defined by the monotonic decreasing solvus reported in Ref. [17] and extends to much lower temperatures than indicated in Ref. [18].

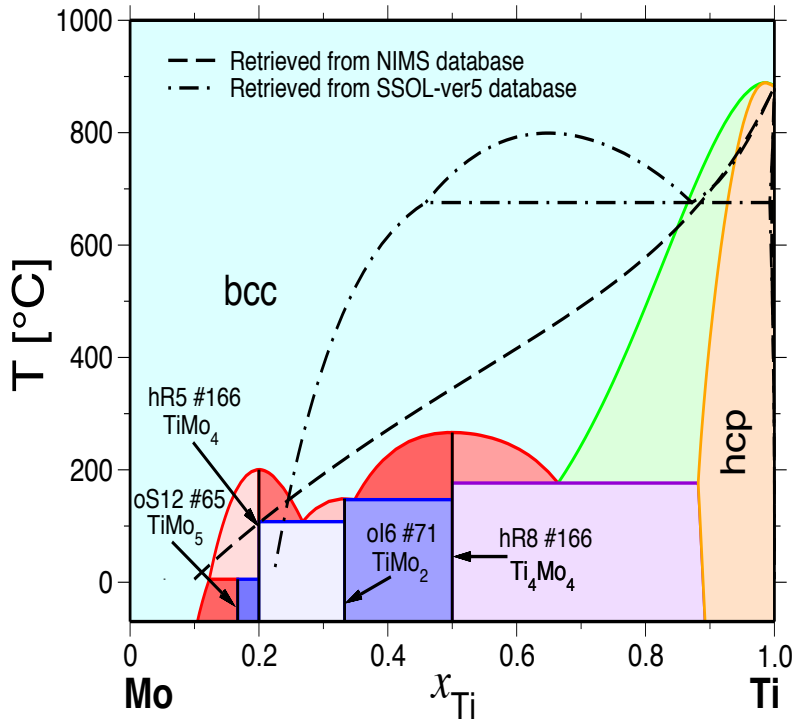

Fig. 8. The calculated Mo-Ti phase diagram based on the Gibbs energy curves of Fig. 7. The dashed lines correspond to the experimental phase diagrams reported in Refs. [17] and [18].

## 4. Conclusion

Accurate description of phase diagrams is fundamental in materials science. It is necessary for designing thermal annealing processes and prediction of the aging processes which may affect alloy homogeneity and consequently its local mechanical properties and corrosion resistance. Two conflicting descriptions of the phase stability of the Mo-Ti *β*-phase have been presented in the literature. One describes a monotonic decreasing *β transus* with increasing Mo content, while the other has a complex *solvus* with a monotectoid *β*-phase separation and a miscibility gap between two different *β* phases. The experimental investigation of the exact location of the *solvus* is difficult since it is located at relatively low temperatures where diffusion is very slow. This ambiguity may also be associated with contaminations or inadequate experimental quench rates.

To provide better estimates of the *solvus* location and structure, we used *ab-initio* calculations to investigate the Mo-Ti phase diagram. We utilize DFT calculations to compute the ground state enthalpy for various ordered structures and for the SQS that mimic the bcc and hcp solid solutions. Negative formation enthalpies obtained for many ordered structures show that stoichiometric compounds are expected in the Mo-Ti system. The predicted compounds become stable at relatively low temperatures and have therefore been difficult to detect by experiments. The excess enthalpy computations for the hcp and bcc solid solutions show attraction interactions between Mo and Ti for the entire concentration range with very weak temperature dependence. The phase diagram constructed from these energies extends the stability domain of the *β*-phase down to the temperatures where the ordered compounds emerge, eliminating the wide bcc-hcp phase separation gap that dominates the experimental phase diagrams at low temperatures.

The new computed phase diagram improves our understanding of the thermodynamics of the system and provides a more solid basis for both fundamental and applied research into Mo-Ti alloys. It should motivate the design of careful experiments to characterize its stable compounds.

An accurate description of its features should be particularly useful for this system, facilitating the development of alloys fulfilling the modern guideline specifications [12] for surgical implant applications.

## Acknowledgements

S. C. and C. T. acknowledge partial support by ONR-MURI under Contract N00014-13-1-0635, DOD-ONR
(N00014-14-1-0526), and the Duke University Center for Materials Genomics. O. L. thanks the Duke University Center for Materials Genomics for its hospitality. The authors acknowledge the CRAY Corporation for computational assistance.